\documentclass[11pt,a4wide]{article}

\usepackage{hyperref}
\usepackage{amsmath}
\usepackage{amsthm}
\usepackage{amsfonts}
\usepackage{amsbsy}
\usepackage{epsfig}
\usepackage{enumitem}
\usepackage{latexsym}
\usepackage{xcolor}
\input amssym.def
\input amssym.tex

\advance \topmargin by -\headheight
\advance \topmargin by -\headsep
\evensidemargin \oddsidemargin
\advance\hoffset by -3mm  
\advance\voffset by  8mm  
\def\bbbc{{\mathchoice {\setbox0=\hbox{$\displaystyle\rm C$}\hbox{\hbox
to0pt{\kern0.4\wd0\vrule height0.9\ht0\hss}\box0}}
{\setbox0=\hbox{$\textstyle\rm C$}\hbox{\hbox
to0pt{\kern0.4\wd0\vrule height0.9\ht0\hss}\box0}}
{\setbox0=\hbox{$\scriptstyle\rm C$}\hbox{\hbox
to0pt{\kern0.4\wd0\vrule height0.9\ht0\hss}\box0}}
{\setbox0=\hbox{$\scriptscriptstyle\rm C$}\hbox{\hbox
to0pt{\kern0.4\wd0\vrule height0.9\ht0\hss}\box0}}}}

\makeatletter
\@addtoreset{equation}{section}
\makeatother

\begin{document}
\title{{\Large \bf{Generating Spherically Symmetric Static Anisotropic Fluid Solutions of Einstein's Equations from Hydrostatic Equilibrium}}}
\author{
M M Akbar$^{}$\footnote{E-mail: akbar@utdallas.edu}
\ \,\,\&\,
R Solanki$^{}$\footnote{E-mail: rahulkumar.solanki@utdallas.edu}
\\
\\ {$^{*}${Department of Mathematical Sciences}}
\\$^{\dagger}${Department of Physics}
\\ {University of Texas at Dallas}
\\ {Richardson, TX 75080, USA}}
\date{\today}
 \maketitle
\begin{abstract}
\noindent
For static fluid spheres, the condition of hydrostatic equilibrium is given by the generalized Tolman--Oppenheimer--Volkoff (TOV) equation, a Riccati equation in the radial pressure. For a perfect fluid source, it is known that finding a new solution from an existing solution requires solving a Bernoulli equation, if the density profile is kept the same. In this paper, we consider maps between static (an)isotropic fluid spheres with the same (arbitrary) density profile  and present solution-generating techniques to find new solutions from existing ones. The maps, in general, require solving an associated Riccati equation, which, unlike the Bernoulli equation, cannot be solved by quadrature. In any case, it can be shown that the output solution is not, in general, regular for a given regular input solution. However, if pressure anisotropy is kept the same, the new solution is both regular and can be found by solving a Bernoulli equation. We give a few examples where the generalized TOV equation, under algebraic constraints, can be converted into a Bernoulli equation and thus, solved exactly. We discuss the physical significance of these Bernoulli equations. Since the density profile remains the same in our approach, the spatial line element is identical for all solutions, which facilitates direct comparison between various equilibrium configurations using fluid variables as functions of the radial coordinate. Finally, combining with the previous study on generation algorithms, we show how this study leads us to a new three-parameter family of exact solutions that satisfy all desirable physical conditions.
\end{abstract}

\section{Introduction}
The spherically symmetric vacuum solution of Einstein's field equations is static, asymptotically flat, and unique (Birkhoff's theorem). However, both static and non-static fluid solutions are possible with spherical symmetry. In relativistic or compact stars, for example, where radiation or degeneracy pressure produces hydrostatic equilibrium against gravity, the spherically symmetric spacetime is static, whereas in the spherical collapse of ``dust'' (i.e., a pressureless fluid), the spacetime is time-dependent. The same pressureless dust, however, can give rise to static configurations under rotation, as in star clusters \cite{misner2017gravitation}.

In this paper, we consider fluid solutions with spherically symmetry, or ``fluid spheres,'' that are static. The spherical symmetry and the existence of a hypersurface-orthogonal Killing vector field imply anisotropy (i.e., the radial and tangential pressures are unequal) and the absence of dissipative effects for the most general fluid configuration; the isotropic limit is a perfect fluid. Physically, anisotropy can occur for various reasons \cite{Herrera:1997plx}. Lema\^{i}tre's solid matter solution (of uniform density) \cite{lemaitre} and Einstein's cluster solution (of arbitrary density profile) \cite{einstein1939stationary}, for both of which the radial pressure vanishes, are probably the earliest (and most extreme) examples of static anisotropic fluid solutions. In Lema\^{i}tre's solution, solid matter is supported against its own gravity by purely transverse stresses. On the other hand, in an Einstein cluster, rotation (due to the circular orbits of the constituent stars) is responsible for the hydrostatic equilibrium. Recently, Einstein's cluster configuration has been applied to model galactic dark matter halos to explain the observed galactic rotation curves (see, e.g., \cite{Lake:2006pp,boehmer2007einstein}). At high densities, the possible existence of a solid core, superfluidity, and superconductivity inside a neutron star could give rise to local anisotropy \cite{Kippenhahn:2012wva}.
Exotic objects, such as boson stars and generalized classes of ``gravastar,'' are intrinsically anisotropic in pressure (see \cite{viaggiu2009modeling} and references therein). In general, for a mixture of two perfect fluids in relative motion, the total energy--momentum tensor in the comoving frame of either is anisotropic in pressure \cite{Letelier:1980mxb}.

Static perfect fluid spheres have been studied extensively from different perspectives \cite{delgaty1998physical,finch1989painleve,kramer1980exact,berger1987general,PhysRevD.67.104015}. The interest in static anisotropic fluid spheres was revived (see e.g., \cite{Rago,herrera2008all,Harko:2002db} and references therein) after the work of \cite{Bowers:1974tgi}. In our previous paper \cite{Akbar:2020qrr}, we presented algorithms that generate all anisotropic solutions and sub-algorithms that generate all regular solutions using the basic functions of the system. We were also able to obtain examples that incorporate other desired physical conditions into regular solutions.
Complementing the study of algorithms, significant insights into the space of solutions can be achieved by studying mappings between solutions, which often leads to simpler algorithms for generating new solutions from old ones. Maps between perfect fluid solutions using spacetime geometry were discussed in \cite{Boonserm:2005ni}. In \cite{boonserm2007solution}, several solution-generating theorems were developed from the condition of hydrostatic equilibrium of perfect fluid spheres; in particular, new solutions were generated from known solutions with the same density profile (see Section~\ref{IsoToIso} for more details). Also, there is an algorithm for generating (general relativistic) anisotropic solutions from isotropic Newtonian solutions \cite{lake2009generating}. In this paper, we study the most general form of the mapping from (an)isotropic to (an)isotropic solutions. Our work can, thus, be seen as natural extensions of these works.

For static (an)isotropic fluid spheres, the condition for hydrostatic equilibrium is  given by the (generalized) Tolman--Oppenheimer--Volkoff (TOV) equation. This is a Riccati equation (even in the isotropic limit), a first-order nonlinear ordinary differential equation in the radial pressure, for which no general solution is known. In the first part of the paper, we find some special conditions under which this equation can be turned into a Bernoulli equation, and thus a general solution can be obtained. We then consider general maps from (an)isotropic solutions to (an)isotropic solutions, i.e., all possible forms of mapping---anisotropic to anisotropic, anisotropic to isotropic, isotropic to isotropic, and isotropic to anisotropic---while keeping density as a function of the radial coordinate invariant.

However, unlike generating isotropic solutions from other isotropic solutions \cite{boonserm2007solution}, with the freedom coming from two different pressure functions, there is a great deal of arbitrariness in generating (an)isotropic solutions from (an)isotropic solutions. First, the generating function can be arbitrarily chosen (subject to basic requirements, e.g., non-negative pressure) to generate (local) anisotropic solutions. However, if the new/output anisotropic solution is required to have a specific form (e.g., a particular tangential pressure), then the question is nontrivial. The generating function then satisfies a (different) Riccati equation (or a Bernoulli equation) and the new solution is not, in general, regular at the center for a given regular input solution. However, we find that when the pressure anisotropy is kept invariant, then the map generates a regular new solution from a regular input solution and the generating function satisfies a Bernoulli equation.
Combining the work in \cite{Akbar:2020qrr} with the study here, we are able to generate a three-parameter exact solution that satisfy all other desirable physical conditions. This shows the complementarity of the work in \cite{Akbar:2020qrr} and the present study.

Throughout the paper, a prime denotes differentiation with respect to the radial coordinate and $a_i$ and $b_i$ (where $ i=1,2,\dots$) are integration constants (or parameters). Units are chosen such that $c=G=1$.

\section{The Spherically Symmetric Static System \label{2}}

The derivation of Einstein equations describing the spherically symmetric static system in canonical coordinates is well known (see, for example, \cite{Bowers:1974tgi, Akbar:2020qrr}). In summary, the system can be described by the metric in canonical/Schwarzschild coordinates:
\begin{equation}
\label{line0}
ds^2= -e^{2\Phi(r)} dt^2+ e^{2\Psi(r)} dr^2 +r^2 (d\theta^2+\sin^2{\theta} d\phi^2),
\end{equation}
with three (Einstein) equations or, equivalently, by the
simplified line element (``curvature coordinates''):
\begin{equation}
\label{line1}
ds^2= -e^{2\Phi(r)} dt^2+ \frac{dr^2}{1-2m(r)/r} +r^2 (d\theta^2+\sin^2{\theta} d\phi^2),
\end{equation}
with one of the two (remaining) Einstein's equations $G^{r}_{r}=8\pi T^{r}_{r}$:
\begin{equation}
\label{EFE}
\frac{d\Phi}{dr}=\frac{m+4\pi r^3 p}{r(r-2m)},
\end{equation}
and the conservation equation, $\nabla_{\mu}T^{\mu}_{r}=0$:
\begin{equation}
\label{momentum_conservation}
P=\frac{r}{2}\left(\frac{dp}{dr}+(\rho+p)\frac{d\Phi}{dr}\right)+p.
\end{equation}
Here, $\Phi(r)$, $\rho(r)$, $p(r)$, and $P(r)$ are the potential, density, radial pressure, and tangential pressure functions, respectively, and
\begin{equation}
\label{density}
m(r)=\int^r_04\pi \rho(x)x^2 dx, \qquad \frac{dm}{dr}=4\pi \rho(r)r^2,
\end{equation}
is the mass function. By substituting equation (\ref{EFE}) into (\ref{momentum_conservation}), one gets the condition for hydrostatic equilibrium, the generalized TOV equation:
\begin{equation}
\label{generalized_TOV}
   -\frac{dp}{dr}+ \frac{2}{r}(P-p)=\left[ \frac{m+4\pi\rho r^3}{r(r-2m)}\right]p+\left(\frac{4\pi r^2}{r-2m}\right)p^2+\frac{\rho m}{r(r-2m)}.
\end{equation}
The generalized TOV equation, written in the slightly peculiar form above, will be central in our discussion below and we will be using this equation with (\ref{EFE}) in the rest of the paper. The perfect/isotropic fluid system can be obtained using $p(r)=P(r)$. Equation (\ref{generalized_TOV}) is then referred to as the TOV equation.\footnote{For the perfect fluid system in isotropic coordinates, see \cite{Glass1978}.} As discussed in \cite{Akbar:2020qrr}, either TOV equation can be viewed as a differential equation in $p(r)$, $m(r)$, and $\Phi'(r)$. For the anisotropic system, this can also be viewed as an algebraic equation defining $P(r)$ in terms of other fluid variables. All these will be important in the rest of the paper.

The radial and tangential pressures can be split into isotropic (trace) and anisotropic (trace-free) parts of the pressure, $\mathbb{P}(r)$ and $\chi(r)$, respectively:
\begin{equation}
\label{EM_splitting}
\mathbb{P}=\frac{(p+2P)}{3}, \qquad \chi=\frac{(P-p)}{3}.
\end{equation}
We will refer to $\chi(r)$ as the pressure anisotropy function.

Physically, equation (\ref{generalized_TOV}) can be seen as the radial-pressure gradient and pressure anisotropy (if positive) supporting against the self-gravitating nonlinear terms on the right (assuming, (i) the density and the radial pressure are monotonically decreasing outwards and (ii) the density and both pressures are positive). Thus, for example, for two equilibrium configurations with the same density profile, if their radial pressures have the same value at some coordinate point but the tangential pressures are different there, the configuration with the lower tangential pressure would have a steeper pressure gradient at that point.

\subsection*{Physical conditions}
There are a number of desirable conditions one often seeks in modeling realistic physical static fluid spheres: regularity at the center, non-negative and monotonically decreasing density and pressures, vanishing of the radial pressure at the boundary, matching with the Schwarzschild (or any other desired) solution at the boundary,\footnote{The continuity and differentiability of the metric components across the boundary (at $r=r_b$) require that
\begin{equation}
  M=m(r_b),\quad e^{2\Phi(r_b)}=1-\frac{2M}{r_b}, \quad p(r_b)=\rho(r_b)=0.
\end{equation}
Physically, the continuity of $\rho(r)$ across the boundary is not required. Fluid spheres satisfying the condition $\rho(r_b)=0$ are called gaseous spheres; see, e.g., \cite{buchdahl1967general}.} subluminal sound speed ($0\leq dp/d\rho\leq 1$ and $0\leq dP/d\rho \leq 1$), and energy conditions ($\rho+p+2P\geq 0$ and $\rho\geq p+2P$).
See, for example, \cite{delgaty1998physical,Harko:2002db,Akbar:2020qrr}.

In \cite{Akbar:2020qrr}, we showed that the three combinations of functions---$(\rho(r)$, $p(r)$), $(\rho(r),\Phi'(r))$, and $(p(r), \Phi'(r))$---lead to an algorithm that can generate all solutions of the system. Conditions for regular solutions (i.e., geometric conditions of regularity when Einstein's equations of the system hold) also turn out to be (equivalent) initial conditions on exactly the same three pairs of variables. Thus all regular anisotropic fluid spheres can be generated using this algorithm directly. Particular solutions that satisfy other physical conditions---in addition to central regularity---can then be attempted and examples were found. Our approach in this paper is similar: our primary objective is general maps between solutions, and we will treat regularity as the first desirable physical condition before embarking upon finding solutions that satisfy all other physical conditions. As we will see, this process does lead us to a  family of exact anisotropic fluid solutions  that satisfy all physical conditions mentioned above as well as the additional condition of decreasing sound speed as a function of the radial coordinate \cite{delgaty1998physical}.

\section{Hydrostatic Equilibrium as a Bernoulli Equation \label{3}}
Equation (\ref{generalized_TOV}) is a Riccati equation in the radial pressure:
\begin{equation}
  \label{Riccati}
  \frac{dp}{dr}+f(r)p+g(r)p^2=h(r)
\end{equation}
where
\begin{equation}
  \label{Riccati_Coeff_anisotropic}
  f(r) = \frac{m+4\pi\rho r^3}{r(r-2m)} + \frac{2}{r},\quad g(r)=\frac{4\pi r^2}{r-2m},\quad h(r)=\frac{2P}{r}-\frac{\rho m}{r(r-2m)}.
\end{equation}
For arbitrary coefficients, the Riccati equation is not solvable by quadrature, but given a particular solution, one can obtain the general solution (see discussions in \cite{boonserm2007solution, Akbar:2020qrr}).

The first question we explore is whether, for a given condition, equation
(\ref{Riccati}) can be turned into a Bernoulli equation:
\begin{equation}
\label{Bernoulli}
\frac{dy}{dr}+A(r)y+B(r)y^2=0,
\end{equation}
which admits solution by quadrature for arbitrary (variable) coefficients.
We discuss five interesting cases in Sections~\ref{Einstein_cluster},~\ref{Bernoulli_inverse},~\ref{Bernoulli_in_anisotrpic_pressure},~\ref{Bernoulli_Herrera}, and~\ref{iso_to_aniso}, when this is possible.
For the first and third cases, there is a natural physical interpretation too.
We choose an algebraic constraint in each case to turn the Riccati equation into a Bernoulli. To determine the coefficients, we still need to specify the density function (before solving it). Note that the trivial $y(r)=0$ solution of (\ref{Bernoulli}) is not necessarily trivial in terms of the fluid variables.

\subsection{Bernoulli in $p(r)$}
\label{Einstein_cluster}
Under the algebraic condition
\begin{equation}
\label{transformation_one_input}
P(r)=\frac{\rho m}{2(r-2m)},
\end{equation}
equation (\ref{generalized_TOV}) turns into the Bernoulli equation (\ref{Bernoulli}) in $p(r)$ with $A(r)=f(r)$ and $B(r)=g(r)$.
It is easy to see that if $p$ is zero at any point, it is zero everywhere, i.e., $p(r)$ vanishes identically. On the other hand, if $dp/dr$ vanishes at some point, $p(r)$ is not necessarily zero. If the coefficients $f(r)$ and $g(r)$ are zero at some point, then $dp/dr$ is also zero at that point with $p$ still nonzero.
If $p(r)\neq 0$, then the general solution is (see, for  example, \cite{zwillinger1998handbook}):
\begin{equation}
\label{solution_1}
p(r)=\frac{a_1\,\exp\left({-\int f(r)dr}\right)}{\left[{1+a_1\int g(r)\exp\left({-\int f(r)dr}\right) dr}\right]}.
\end{equation}
Using (\ref{density}) and some standard integration techniques, we obtain:
\begin{equation}
\label{solution_2}
p(r)=\frac{a_1\,\exp(\Psi)}{r^2\left[1+4\pi a_1\int \exp(3\Psi)/r \, {dr} \right]}.
\end{equation}
From equations (\ref{EFE}) and (\ref{solution_2}), upon integration, we obtain:
\begin{equation}
\label{solution_3}
e^{2\Phi(r)}=\left({1+4\pi a_1\int \exp(3\Psi) \frac{dr}{r}} \right)^2\exp\left(\int \frac{2mdr}{r(r-2m)}\right),
\end{equation}
in which the multiplicative constant of integration is absorbed into the time coordinate. Therefore, from one input $\rho(r)$, we can generate anisotropic solutions using equations (\ref{transformation_one_input}), (\ref{solution_2}), and (\ref{solution_3}).

For $p(r)=0$, the class of anisotropic solutions satisfying condition (\ref{transformation_one_input}) are known as Florides' solutions \cite{florides1974new}.
In sharp contrast to the anisotropic/perfect fluid system, hydrostatic equilibrium in this case results solely from the nonzero tangential pressure.
Physically, this class of solutions represents Einstein clusters.\footnote{In an Einstein cluster, the constituent stars, i.e., dust particles, move in concentric circular spatial orbits
such that there is no preferred axis of rotation.
Each star follows a timelike geodesic of the spacetime produced by all stars together.
Through dynamical equilibrium, matter as a whole is stationary, i.e., the spatial component of the four-velocity is zero. The general approach to modeling a large number of gravitating masses (e.g., a star cluster) as a static fluid sphere
is to apply kinetic theory in curved spacetime (see, e.g., \S 25.7 of MTW \cite{misner2017gravitation} and references therein).}
For a given $\rho(r)$, the rest of the solution is given by setting $a_1=0$ in equation (\ref{solution_3}):
\begin{equation}
\label{solution_4}
e^{2\Phi(r)}=\exp\left(\int \frac{2mdr}{r(r-2m)}\right).
\end{equation}
Thus, there are two fluid configurations with the same density and tangential pressure, which maintain hydrostatic equilibrium with zero and nonzero radial pressures. This shows that the radial pressure in the latter is being counterbalanced by its own gradient (since, in general relativity, pressure gravitates).

For the class of solutions with nonzero radial pressure, the $g_{tt}$ component of the metric is not differentiable across the boundary if matched with a Schwarzschild vacuum solution.
Moreover, since the tangential pressure vanishes at the center for finite $\rho(0)$, the regularity condition of isotropy of the central pressure is not satisfied.\footnote{Choosing, $(P-p)=\rho m/2(r-2m)$ instead, converts (\ref{generalized_TOV}) into a Bernoulli equation in $p(r)$ without the factor $2/r$ (see \textbf{Lemma 2} below).
These regular solutions, however, cannot be matched with the Schwarzschild exterior, since $p(r)$ is nonzero everywhere.
In \cite{viaggiu2009modeling}, the author generates solutions with nonzero radial pressure from a constant-density Florides' solution. These solutions are regular and can be matched with the Schwarzschild exterior.
However, of course, the conditions (\ref{transformation_one_input}) and $(P-p)=\rho m/2(r-2m)$ are not satisfied by the generated solutions.}

\subsection{Bernoulli in $1/p(r)$}
\label{Bernoulli_inverse}
The condition
\begin{equation}
\label{EOS}
p(r)=-\rho(r)
\end{equation}
renders (\ref{momentum_conservation}) free from the $\Phi'$ term and the generalized TOV equation reduces to
\begin{equation}
\label{simplifiedTOV}
\frac{dp}{dr}+\frac{2}{r}p=\frac{2}{r}P.
\end{equation}
It is a Bernoulli in $z(r)=1/p(r)$.
Moreover, from (\ref{EFE}) and (\ref{EOS}), we get:
\begin{equation}
\label{solution_5}
e^{2\Phi(r)}=\left(1-\frac{2m(r)}{r}\right)=e^{-2\Psi(r)}.
\end{equation}
In the isotropic limit, $P(r)=p(r)$, the choice $p(r)=-\rho(r)$ would necessarily render $p(r)$ a constant function, which is the well-known case of the cosmological constant.
However, in the anisotropic case, there is a class of solutions satisfying the relation (\ref{solution_5}).

At the center, if the density is finite, then from condition (\ref{EOS}), the radial pressure is also finite.
Thus, all one needs for a regular solution is to choose a density profile such that $\rho(0)$ is finite, and integrate $\rho(r)$ only once to obtain $m(r)$.

\subsection{Bernoulli in $\chi(r)$}
\label{Bernoulli_in_anisotrpic_pressure}
From equation (\ref{EM_splitting}), we have
\begin{equation}
\label{transformation}
\begin{aligned}
p(r)&=\mathbb{P}(r)-2\chi(r), \\
P(r)&=\mathbb{P}(r)+\chi(r).
\end{aligned}
\end{equation}
If $\mathbb{P}(r)$ and $\rho(r)$ satisfy the TOV equation of the perfect fluid system $\{\rho(r), p_0(r)\equiv \mathbb{P}(r), \Phi_0(r)\}$, then the generalized TOV equation turns into the following Bernoulli equation in $\chi(r)$:
\begin{equation}
\label{bernoulli_chi}
\frac{d\chi}{dr} +\bar{f}(r)\chi+\bar{g}(r)\chi^2=0,
\end{equation}
where
\begin{equation}
\label{coefficients-2}
\bar{f}(r)=\frac{3}{r}+\frac{m+4\pi r^3(\rho+2p_{0})}{r(r-2m)}, \quad \bar{g}(r)=-\frac{8\pi r^2}{r-2m}.
\end{equation}
To integrate (\ref{bernoulli_chi}), it is useful to rewrite $\bar{f}(r)$ as
\begin{equation}
\bar{f}(r)=\frac{4}{r}-\frac{d/dr\{r(r-2m)\}}{2r(r-2m)}+2\frac{d\Phi_0}{dr}.
\end{equation}
The general solution is then
\begin{equation}
\label{gamma}
\chi(r)=\frac{a_2\,\exp(-\Psi-2\Phi_0)}{r^3 \left[1-8\pi a_2\int{\exp(\Psi-2\Phi_0)/r^2 \, dr} \right]},
\end{equation}
and the new potential is
\begin{equation}
\label{phi_before}
\Phi(r)=\Phi_0(r)-\int \frac{8\pi r^2 \chi}{r-2m}dr.
\end{equation}
To find $\Phi$, we actually do not need to evaluate the two integrals of (\ref{gamma}) and (\ref{phi_before}) one after another. Applying standard integration techniques, we get\footnote{Reference \cite{Maharaj:2005vb} obtains a similar result with a different approach without using a Bernoulli equation.}
\begin{equation}
\label{phi}
e^{2\Phi}=e^{2\Phi_0} \left(1 -8\pi a_2\int e^{\Psi-2\Phi_0}\frac{dr}{r^2}\right)^2.
\end{equation}
Therefore, for any given perfect fluid solution $\{\rho(r),p_0(r),\Phi_0(r)\}$, one can generate an anisotropic fluid solution $\{\rho(r),p(r),P(r),\Phi(r)\}$ via equations (\ref{transformation}), (\ref{gamma}), and (\ref{phi}).
Since $\chi(r)$ is nonzero everywhere, the central pressure is not isotropic; hence, the output solution is not regular.

Note that, since the change in the radial pressure is compensated for by the tangential pressure, equation (\ref{bernoulli_chi}) indicates how much anisotropy can be introduced without disturbing the hydrostatic equilibrium for any given density profile.
Moreover, if the density and the isotropic part of the pressure are positive and monotonically decreasing---which are desirable in physical models---it follows from (the coefficients of) equation (\ref{bernoulli_chi}) that if $\chi<0$ at some point (i.e., $p>P$) then $\chi'>0$. Further, from equation (\ref{transformation}), one gets after differentiation, $p'<p_0'$ at that point. In other words, if the radial pressure is increased at a point, the tangential pressure decreases there (since the isotropic part of the pressure is invariant). Furthermore, the radial pressure gradient becomes steeper (relative to the isotopic configuration) at that point to make up for the reduced tangential pressure; the converse also holds.\footnote{From equations (\ref{bernoulli_chi}), (\ref{coefficients-2}), and (\ref{transformation}):
\begin{equation}
    \frac{d\chi}{dr}=\chi\left[\frac{8\pi r^3(\chi-p_0)}{r(r-2m)}-\frac{3}{r}-\frac{m+4\pi r^3\rho}{r(r-2m)}\right]=-\chi\left[\frac{8\pi r^3(\chi+p)}{r(r-2m)}+\frac{3}{r}+\frac{m+4\pi r^3\rho}{r(r-2m)}\right].
\end{equation}
Thus, if $\chi>0$ at some point, then $\chi'<0$ at that point.} 

\subsection{Bernoulli from Conformal Flatness}
\label{Bernoulli_Herrera}
The condition of conformal flatness (i.e., the vanishing of the Weyl tensor of (\ref{line0}) and (\ref{line1})) also works out to be an algebraic condition (see \cite{herrera2008all,Herrera:2001vg}):
\begin{equation}
  \label{Weyl_EM}
  P-p=\frac{3m}{4\pi r^3}-\rho.
\end{equation}
The resulting Bernoulli equation is
\begin{equation}
  \label{Bernoulli_Conformally_flat}
  \frac{d\xi}{dr}+\left(\Psi'+\frac{2}{r}-\frac{2e^{\Psi}}{r}\right)\xi+\left(4\pi r e^{2\Psi}\right)\xi^2=0,
\end{equation}
where
\begin{equation}
  \xi(r)=p-\frac{(3e^{-\Psi}+1)(e^{-\Psi}-1)}{8\pi r^2}.
\end{equation}

\section{A General Approach to Mapping \label{4}}
In \cite{Akbar:2020qrr}, we studied algorithms and sub-algorithms that generate all solutions and all regular solutions of the anisotropic system.
We have also shown how one can utilize generating algorithms that have linear equations to create particular regular solutions that can further be used to obtain the general solution of the Riccati equation.
Below we will study the general question of mapping between solutions within and {\emph{between}} the isotropic and anisotropic systems.
This will also extend the work of \cite{Boonserm:2005ni} on the isotropic system.
 \\
 \\
\textbf{Theorem 1:} The solution $\lambda(r)$ of
\begin{equation}
\label{A2A_tangential_pressure}
\frac{d\lambda}{dr} + F(r)\lambda +G(r) \lambda^2=\frac{2}{r}\frac{(P_1-P_2)}{p_1},
\end{equation}
where
\begin{equation}
\label{A2A_coeff}
F(r)=\frac{2}{r}+\Phi'_1+\frac{p'_1}{p_1}+\frac{4\pi r^2(\rho+p_1)}{r-2m},\quad G(r)=-\frac{4\pi r^2p_1}{r-2m}
\end{equation}
maps an anisotropic solution $\{\rho,p_1,P_1,\Phi_1\}$ into another anisotropic solution $\{\rho,p_2\equiv (1-\lambda)p_1,P_2,\Phi_2\equiv \Phi_1-\int {4\pi r^2 p_1 \lambda}/{(r-2m)}dr\}$. In addition,
\begin{enumerate}
  [label=\normalfont(\alph*)]
  \item The inverse map is given by (for $\lambda(r)\neq 1$ and $p_1(r)\neq 0$):
  \begin{equation}
    \label{A2A_inverse}
    \lambda^{-1}(r)=\frac{\lambda}{\lambda-1}.
  \end{equation}
  \item Successive maps by $\lambda_i$'s $(i=1,2,\dots,n-1)$ taking an anisotropic solution $\{\rho,p_1,P_1,\Phi_1\}$ into the final solution $\{\rho,p_n,P_n,\Phi_n\}$ is independent of the order chosen and equivalent to a map given by:
  \begin{equation}
    \label{composite}
    \Lambda=1-\prod_{i=1}^{n-1}(1-\lambda_i).
  \end{equation}
\end{enumerate}
\begin{proof}
Since both $\{\rho,p_1,P_1,\Phi_1\}$ and  $\{\rho,p_2,P_2,\Phi_2\}$
satisfy equations (\ref{EFE}) and (\ref{momentum_conservation}), substituting the relation $p_2=(1-\lambda)p_1$ into them gives
\begin{equation}
\label{A2A_potential}
     \Phi_2=\Phi_1-\int \frac{4\pi r^2 p_1 \lambda}{r-2m}dr
\end{equation}
and (\ref{A2A_tangential_pressure}), respectively.
In other words, equation (\ref{A2A_tangential_pressure}) is just a restatement that both solutions belong to the same system.\footnote{Here, the generating function
\begin{equation}
  \label{A2A_transformation}
  \lambda = \frac{p_1-p_2}{p_1},
\end{equation}
represents the fractional change in the radial pressure, from one solution to the other. For $\lambda(r)=1$, the radial pressure $p_2(r)$ vanishes and we obtain Florides' solutions.  On the other hand, if the known solution is chosen as Florides' class (i.e., $p_1(r)=0$), then the generating function maps to the same solution. This can be seen as the null function of the mapping.}
\begin{enumerate}
  [label=(\alph*)]
  \item From equation (\ref{A2A_inverse}), $(1-\lambda)=1/(1-\lambda^{-1})$. Substituting this relation into $p_2=(1-\lambda)p_1$ gives $p_1=(1-\lambda^{-1})p_2$. Furthermore, substituting both $\lambda$ and $p_1$ into (\ref{A2A_potential}) and (\ref{A2A_tangential_pressure}), upon simplification, gives
  \begin{align}
    \Phi_1 &= \Phi_2-\int \frac{4\pi r^2 p_2 \lambda^{-1}}{r-2m}dr, \\
    P_1    &= P_2-\frac{rp_2}{2}\left[\frac{d\lambda}{dr}^{-1} + F_2(r)\lambda^{-1} +G_2(r) (\lambda^{-1})^2 \right],
  \end{align}
where $F_2(r)$ and $G_2(r)$ are the same as in (\ref{A2A_coeff}) except $\Phi_2$ and $p_2$ replace $\Phi_1$ and $p_1$. Thus, $\lambda^{-1}(r)$ is the generating function of the inverse map.
  \item The successive application of $\lambda_i$'s, from the definition of the generating function, yields
  \begin{equation}
    \label{p_n}
    p_n=(1-\lambda_{n-1})p_{n-1}=p_1\prod_{i=1}^{n-1}(1-\lambda_i)=(1-\Lambda)p_1,
  \end{equation}
which clearly shows the commutative nature of the $\lambda_i$'s. The resulting $\Phi_n$ and $P_n$ are likewise independent of the order of $\lambda_i$'s, as can be seen from their integral expressions:
  \begin{align}
    \Phi_n &= \int \frac{m\,dr}{r(r-2m)}+\int \frac{4\pi r^2p_1}{r-2m}\left(\prod_{i=1}^{n-1}(1-\lambda_i)\right)dr \equiv \Phi_1-\int \frac{4\pi r^2 p_1 \Lambda}{r-2m}dr,\\
    P_n    &= P_1-\frac{rp_1}{2}\left[\frac{d\Lambda}{dr} + F(r)\Lambda +G(r) \Lambda^2 \right].
  \end{align}\qedhere
  \end{enumerate}
\end{proof}
Note that the above theorem also implies that if one divides the solution space of the anisotropic system into equivalent density classes, then within each density class, all solutions are ``generated'' algebraically from any solution via all possible functional forms of $\lambda(r)$ using $p_2=p_1(1-\lambda)$. One only needs to solve for $\lambda(r)$ using (\ref{A2A_tangential_pressure}) when information about $p_2(r)$ is not known in advance in some form or other. In the former case, if $\rho(0)$ and $p_1(0)$ are finite, then choosing finite $\lambda(0)$ would ensure regularity; this is essentially Theorem 4.5 of \cite{Akbar:2020qrr}.
In the isotropic case (i.e., $p(r)=P(r)$), $\lambda(r)$ can only be solutions of Bernoulli equations.
\subsubsection*{Special cases}
This Riccati equation (\ref{A2A_tangential_pressure}) turns into a Bernoulli in $\lambda(r)$ when
\begin{enumerate}
  [label=(\roman*)]
  \item The tangential pressure $P(r)$ is kept invariant:
  \begin{equation}
    \label{tangential}
    \frac{d\lambda}{dr} +F(r)\lambda +G(r) \lambda^2=0.
  \end{equation}
  \item The isotropic part of the pressure $\mathbb{P}(r)$ is kept invariant:
  \begin{equation}
    \label{isotropic}
    \frac{d\lambda}{dr} + \left(F(r)+\frac{1}{r}\right)\lambda +G(r) \lambda^2=0.
  \end{equation}
  \item The anisotropic part of the pressure $\chi(r)$ is kept invariant:
  \begin{equation}
    \label{anisotropic}
    \frac{d\lambda}{dr} + \left(F(r)-\frac{2}{r}\right)\lambda +G(r) \lambda^2=0.
  \end{equation}
\end{enumerate}
As a final special case, note that if $p_2(r)=p_1(r)$, then $\lambda(r)=0$, $\Phi_2=\Phi_1$, and $P_2=P_1$. Physically, this means that the gradient of the radial pressure cannot be changed to support any change in the tangential pressure (under the condition that the radial pressure is not changed).
\\
\\
\textbf{Lemma 2:}  From a known anisotropic solution $\{\rho,p_1,P_1,\Phi_1\}$, a new anisotropic solution $\{\rho,p_2,P_2,\Phi_2\}$ can be generated by keeping the density and pressure anisotropy invariant, and
\begin{equation}
  \label{regular_radial}
  p_2=p_1-\frac{b_1\,e^{-\Psi-2\Phi_1}}{\left[1-b_1\int 4\pi re^{\Psi-2\Phi_1}dr \right]},
\end{equation}
\begin{equation}
  \label{regular_tangential}
  P_2=P_1-\frac{b_1\,e^{-\Psi-2\Phi_1}}{\left(1-b_1\int 4\pi re^{\Psi-2\Phi_1}dr \right)},
\end{equation}
\begin{equation}
  \label{regular_phi}
  e^{2\Phi_2}=e^{2\Phi_1} \left(1-b_1\int 4\pi re^{\Psi-2\Phi_1}dr\right)^2.
\end{equation}
In addition, the new solution is regular at the center if the input solution is.
Furthermore, if the pressure gradient at the center vanishes for the known solution, then it vanishes for the new solution.
\begin{proof}
As noted earlier, the Riccati equation (\ref{A2A_tangential_pressure}) becomes a Bernoulli equation in $\lambda(r)$, if $P_2=P_1$ (or $\mathbb{P}_2=\mathbb{P}_1$). However, the term $2/r$ (or $3/r$) within the coefficient of $\lambda(r)$ precludes regularity at the center. The choice of
\begin{equation}
  \label{A2A_regular}
  P_2=P_1-\lambda p_1,
\end{equation}
instead renders the coefficient of $\lambda(r)$ free of the $2/r$ term.
The condition (\ref{A2A_regular}), simply, is $\chi_2=\chi_1$, due to the definition of the generating function (\ref{A2A_transformation}). For $\lambda(r)\neq 0$, the solution of equation (\ref{anisotropic}) is
\begin{equation}
  \label{regular_Bernoulli}
  \lambda(r)=\frac{b_1\,e^{-\Psi-2\Phi_1}}{p_1\left[1-b_1\int 4\pi re^{\Psi-2\Phi_1}dr \right]}.
\end{equation}
Substituting this into equations (\ref{A2A_transformation}), (\ref{A2A_regular}), and (\ref{A2A_potential}) gives the rest of the solution.

For a regular solution, $\rho(0)$ and $p_1(0)$ are finite and as $r\to 0$, the coefficients of the Bernoulli equation (\ref{anisotropic}) vanish. Moreover, from the solution (\ref{regular_Bernoulli}), $\lim_{r\to 0} \lambda(r)$ is finite. Therefore, the new solution is regular, since $\rho(0)$ and $p_2(0)=P_2(0)=p_1(0)[1-\lambda(0)]$ are finite. Furthermore, $\lambda'(0)=0$, since $\lambda(0)$ is finite and the coefficients of the Bernoulli equation (\ref{anisotropic}) vanish at the center. Therefore, from equation (\ref{A2A_transformation}), after differentiation, $p'_2(0)=0$ if $p'_1(0)=0$.
\end{proof}

Applying \textbf{Lemma 2} to a one-parameter family of solutions, for example, with identical density and tangential pressure as the Einstein cluster (Section~\ref{Einstein_cluster}), would yield a two-parameter family of solutions.
These solutions are regular if the parameter $a_1=0$ in (\ref{solution_3}).
In other words, the regular solutions are still a one-parameter family.
Since the general solution of the TOV equation of a perfect fluid is a one-parameter family, applying \textbf{Lemma 2} (with $\chi(r)=0$) to a particular solution gives the general solution.

\subsection{Isotropic to Anisotropic Mapping}
\label{iso_to_aniso}
If $p_1(r)=P_1(r)$, then the map generates an anisotropic solution from an isotropic solution. The tangential pressure of the new solution is given by
\begin{equation}
\label{I2A_tangential_pressure}
P_2=p_1 \left[1-\frac{r}{2} \left( \frac{d\lambda}{dr} + F(r)\lambda +G(r) \lambda^2\right) \right].
\end{equation}
As noted earlier, if the radial pressure is kept invariant, then $\lambda(r)=0$, $\Phi_2(r)=\Phi_1(r)$, and $P_2(r)=p_1(r)$.
Thus, for perfect fluid solutions, anisotropy cannot be introduced if the radial pressure (and density) remains the same.

If a specific form of tangential pressure is required, then we can solve for $\lambda(r)$ using equation (\ref{I2A_tangential_pressure}), which is a Riccati equation.
For example, if $P_2(r)=p_1(r)$, then equation (\ref{I2A_tangential_pressure}) becomes a Bernoulli (\ref{tangential}).
For $\lambda(r)\neq 0$, the solution is
\begin{equation}
\label{I2A_Bernoulli}
\lambda(r)=\frac{b_2\,e^{-\Psi-2\Phi_1}}{r^2p_1\left[1-b_2\int \frac{4\pi}{r}e^{\Psi-2\Phi_1}dr \right]}.
\end{equation}
Thus, the solution reduces to evaluating only one integral, other than the mass function.
Finally, from equation (\ref{A2A_transformation}),
\begin{equation}
\label{sol_I2A_tangential_pressure}
P_2-p_2=p_1\lambda=\frac{b_2\,e^{-\Psi-2\Phi_1}}{r^2\left[1-b_2\int \frac{4\pi}{r}e^{\Psi-2\Phi_1}dr \right]}.
\end{equation}
Therefore, unlike the radial pressure function, when the tangential pressure remains the same, anisotropy can be introduced in perfect fluid solutions using equation (\ref{sol_I2A_tangential_pressure}).
However, as noted earlier, the presence of the $2/r$ term in $F(r)$ precludes regularity, even if the perfect fluid solution is regular.

The class of solutions calculated in Section~\ref{Bernoulli_in_anisotrpic_pressure} can be recovered using the generating function $\lambda(r)$. If the isotropic part of the pressure is kept invariant i.e., $\mathbb{P}_2=p_1$, then from equation (\ref{A2A_transformation}),
\begin{equation}
  \frac{2}{r}\left(1-\frac{P_2}{p_1}\right)=-\frac{\lambda}{r},
\end{equation}
and the Riccati equation (\ref{I2A_tangential_pressure}) in $\lambda(r)$ turns into a Bernoulli equation (\ref{isotropic}).
The rest of the solution follows from Section~\ref{Bernoulli_in_anisotrpic_pressure}.

\subsection{Anisotropic to Isotropic Mapping}
With $p_2(r)=P_2(r)$, equation (\ref{A2A_tangential_pressure}) becomes
\begin{equation}
\label{A2I_tangential_pressure}
\frac{d\lambda}{dr}+\left[F(r)-\frac{2}{r}\right]\lambda+G(r)\lambda^2=-\frac{2}{r}\left(1-\frac{P_1}{p_1}\right).
\end{equation}
Here $\lambda(r)$ cannot take an arbitrary value.
This is the same as solving the TOV equation for the isotropic system.

\subsection{Isotropic to Isotropic Mapping \label{IsoToIso}}
For $p_1(r)=P_1(r)$ and $p_2(r)=P_2(r)$, the mapping is between isotropic solutions, and the choice of $\lambda(r)$ is not arbitrary.
This mapping coincides with the first solution-generating theorem given in \cite{boonserm2007solution}, where the authors utilize the following property of the Riccati equation.
If $\{\rho(r),p_1(r)\}$ satisfies the TOV equation and $p_n(r)$ satisfies
\begin{equation}
\label{complimentary_sol_to_Riccati}
\frac{dp_{n}}{dr}+[f_0(r)+2g_0(r)p_1]p_n+g_0(r)p_n^2=0,
\end{equation}
then $\{\rho(r),p_1(r)+p_n(r)\}$ satisfies the TOV equation.
The authors generate a one-parameter family of solutions by integrating the Bernoulli equation (\ref{complimentary_sol_to_Riccati}), and the shift in central pressure $(\delta p_{c})$ is chosen as this parameter, since the particular solution $\{\rho(r),p_1(r)\}$ is considered regular.
Consider a given solution for which the density and pressure are positive and monotonically decreasing with the radial coordinate.
At some point, if $p_n>0$, then from equation (\ref{complimentary_sol_to_Riccati}), $p'_n<0$. In words, if the pressure is increased, the pressure gradient becomes steeper to maintain the hydrostatic equilibrium and vice versa.
This discussion is valid for \textbf{Lemma 2}, except that the radial pressure replaces the isotropic pressure.

\section{Harvesting the Riccati: An Exact Physical Solution}
As we mentioned earlier, the general solution of a Riccati equation can be found only if one can supply a particular solution. In \cite{Akbar:2020qrr}, we have seen that one can use a different algorithm to generate particular solutions that can be used for this purpose. If the first algorithm gives a one-parameter solution, the Riccati will turn it into a two-parameter family. The freedom due to the additional parameter increases the possibility of incorporating more physical conditions. Below we will see that the particular nature of \textbf{Lemma 2} above has an added advantage in this regard.

Here we will combine \textbf{Theorem 4.1} of \cite{Akbar:2020qrr} with \textbf{Lemma 2}.
Consider the following input functions, which satisfy the regularity conditions $\Phi'(0)=0$ and $\Phi''(0)$, $P(0)$ finite:
\begin{equation}
  \label{H_inputs}
  P_1(r)=\frac{-r^4+A^2r^2+B^2(A^2-B^2)}{8\pi\,A^2\,B^2(B^2-r^2)},\quad \Phi_1(r)=-\frac{1}{2}\ln\left(1-\frac{r^2}{B^2}\right),
\end{equation}
in which $0\leq r<|B|$ and $A$ and $B$ are nonzero constants.
The mass function follows from \textbf{Theorem 4.1}:
\begin{equation}
  m_1(r)=\frac{r^3(A^2+B^2-r^2)}{2A^2\,B^2}.
\end{equation}
Moreover, the radial pressure, density, and pressure anisotropy are
\begin{align}
  p_1(r)    &= \frac{A^2-B^2-r^2}{8\pi A^2\,B^2}, \\
  \rho_1(r) &= \frac{3(A^2+B^2)-5r^2}{8\pi A^2\,B^2}, \label{H_density} \\
  \chi_1(r) &= \frac{r^2}{12\pi A^2 B^2}\left(\frac{A^2-r^2}{B^2-r^2}\right).
\end{align}
The resulting anisotropic solution is
\begin{equation}
  \label{H_anisotropic}
  ds^2=-\frac{dt^2}{\left(1-\frac{r^2}{B^2}\right)}+\frac{dr^2}{\left(1-\frac{r^2}{A^2}\right)\left(1-\frac{r^2}{B^2}\right)}+r^2(d\theta^2+\sin^2\theta\,d\phi^2).
\end{equation}
If, instead, $\Phi_1(r)$ and $\chi_1(r)$ are chosen as input functions, then the solution contains an additional parameter, the central density $\rho(0)$.
Then the mass function, from \textbf{Theorem 4.2}, is
\begin{equation}
  m_3(r)=m_1(r)+\frac{r^3\,e^{2r^2/B^2}}{6\,A^2B^2}\left(1-\frac{r^2}{B^2}\right)^3 \left(8\pi A^2 B^2 \rho(0)-3B^2-3A^2\right).
\end{equation}
For the seed solution (\ref{H_inputs})--(\ref{H_anisotropic}), the density and radial pressure are monotonically decreasing functions and the central density is positive for any nonzero $A$ and $B$.
The radial sound speed, $dp/d\rho=1/5$, is subluminal.
The central pressure is positive if $A^2>B^2$ and the boundary at which the radial pressure vanishes is at $r_{b,1}=\sqrt{A^2-B^2}$.
Since $r_{b,1}<|B|$, then $2B^2>A^2$ and the density at the boundary is positive.
Thus, both the density and radial pressure, which are monotonically decreasing functions with non-negative values at the center and at the boundary, are non-negative.
Furthermore, under these conditions, the pressure anisotropy is non-negative; therefore, the tangential pressure is positive.
However, the tangential pressure is monotonically increasing:
\begin{equation}
  8\pi P'_1(r)=\frac{2r}{A^2B^2}+\frac{4r(A^2-B^2)}{A^2(B^2-r^2)^2}.
\end{equation}
The trace energy condition, i.e., $\rho\geq p+2P$ \cite{Martin-Moruno:2017exc}, is still satisfied, since, for $B^2<A^2<2B^2$, $\rho(r)-p(r)$ is monotonically decreasing and $\rho(r_{b,1})>2P(r_{b,1})$.
Thus, the seed solution ({\ref{H_anisotropic}}) satisfies all physical conditions if
\begin{equation}
  B^2<A^2<2B^2,
\end{equation}
where $r\in [0,\sqrt{A^2-B^2}]$, except that the tangential pressure is monotonically increasing, which leads to an imaginary tangential sound speed.
Below we will see how \textbf{Lemma 2} cures this problem.

\subsubsection*{A Family of Exact Physical Anisotropic Fluid Spheres}
Since the anisotropic solution (\ref{H_anisotropic}) is regular, we can rewrite equations (\ref{regular_radial}) and (\ref{regular_tangential}), i.e., $\Delta p=p_2-p_1=P_2-P_1$, in terms of the shift in the central pressure $\Delta p_0$:
\begin{equation}
  \Delta p=\frac{\Delta p_0\, e^{-\Psi_1-2\Phi_1}}{e^{-2\Phi_1(0)}+\Delta p_0\, I(r)}\quad \text{where} \quad I(r)=4\pi\int_{0}^{r} x e^{\Psi_1-2\Phi_1} dx.
\end{equation}
It is clear from the equation above that $\Delta p >0$, if $\Delta p_0 >0$ and the seed metric retains its Lorentzian signature under the change in boundary (i.e., the $r=$ constant and $t=$ constant hypersurfaces remain spacelike and timelike respectively).
Furthermore,
\begin{equation}
  \Delta p'=\frac{\Delta p_0\,\frac{d}{dr}\left(e^{-\Psi_1-2\Phi_1}\right)}{e^{-2\Phi_1(0)}+\Delta p_0\, I(r)}-\frac{4\pi r\, \Delta p_0^2\, e^{-4\Phi_1}}{\left(e^{-2\Phi_1(0)}+\Delta p_0 \, I(r) \right)^2},
\end{equation}
is negative if $\exp(-\Psi_1-2\Phi_1)$ is monotonically decreasing.

\begin{figure}[ht!]
  \makebox[\textwidth][c]{\includegraphics[width=0.8\textwidth]{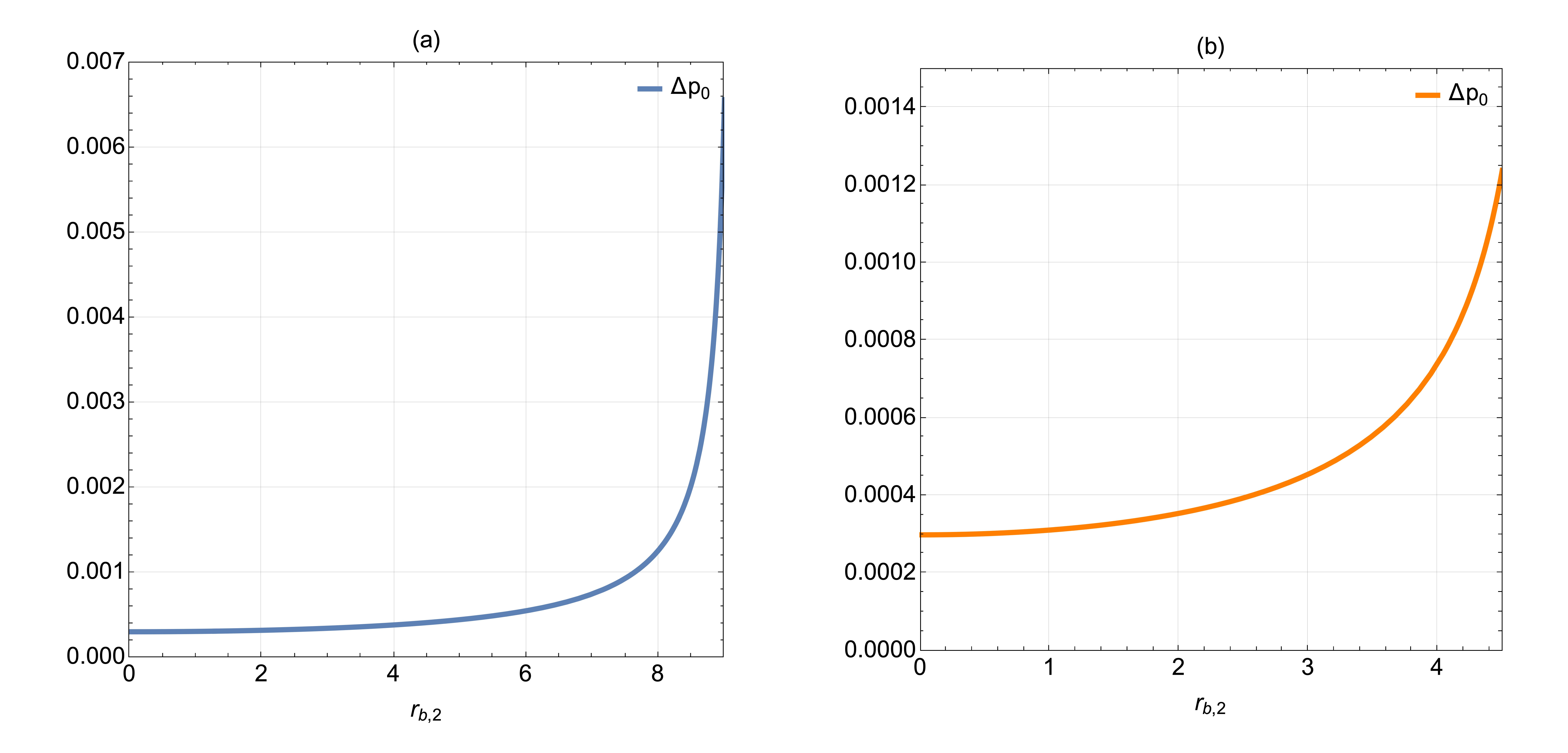}}
  \caption{For (a) $A=10$, $B=20$, and (b) $A=5$, $B=10$, both plots represent a range of the boundary $r_{b,2}$ for which the shift in the central pressure $\Delta p_0$ is positive. Therefore, $\Delta p(r)$ is positive and monotonically decreasing. Furthermore, $\rho(r)$ and $p_2(r)$ in this range are non-negative and monotonically decreasing, and $P_2(r)$ is positive. }
  \label{fig:Range_Second_Paper}
\end{figure}

\begin{figure}[ht!]
  \makebox[\textwidth][c]{\includegraphics[width=1.15\textwidth]{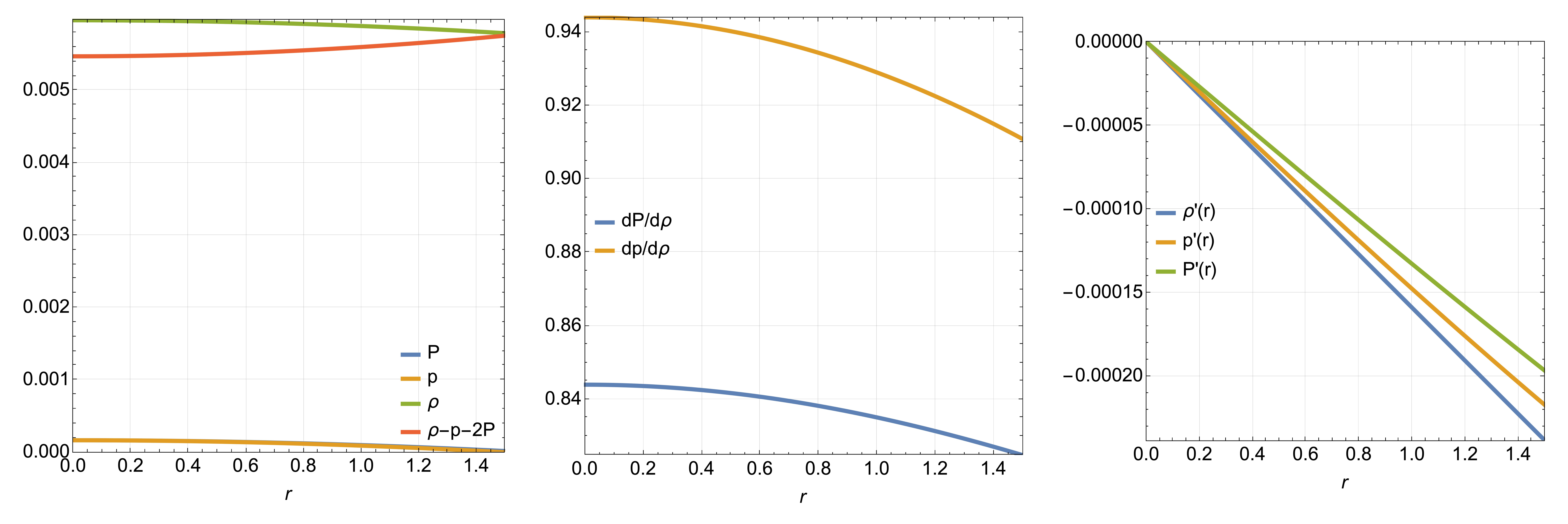}}
  \caption{Here, $A=5$, $B=10$ and $r_{b,2}=1.5$ yield the shift in the central pressure $\Delta p_0=0.000135$. The solution satisfies all the physical conditions. The sound speed is also monotonically decreasing.}
  \label{fig:Example_Second_Paper}
\end{figure}

For the seed metric (\ref{H_anisotropic}), we get
\begin{equation}
\label{Ir}
  I(r)=2\pi A^2\left[1-\sqrt{\left(1-\frac{r^2}{A^2}\right)\left(1-\frac{r^2}{B^2}\right)} + \left(\frac{B^2-A^2}{|AB|}\right)\ln\left(\frac{|A|+|B|}{\sqrt{A^2-r^2}+\sqrt{B^2-r^2}} \right) \right].
\end{equation}
Here, we may choose three parameters, $\Delta p_0>0$, $A$, and $B$, and find the boundary $r_{b,2}$ using the relation $p_2(r_{b,2})=0$ such that $r^2_{b,2}<\min\{A^2,B^2\}$. However, this requires solving a transcendental equation. Instead, we can choose $A,\, B$ and $r^2_{b,2}<\min\{A^2,B^2\}$ and find $\Delta p_0$, from $p_2(r_{b,2})=0$, such that it is positive.
Under these conditions, $\Delta p(r)$ is positive and monotonically decreasing, since
\begin{equation}
  \frac{d}{dr}\left(e^{-\Psi_1-2\Phi_1}\right)=-r\left(\frac{(B^2-r^2)^{3/2}}{\sqrt{A^2-r^2}}-3\,\sqrt{(A^2-r^2)(B^2-r^2)} \right) <0.
\end{equation}
Furthermore, the density remains positive and monotonically decreasing.
If $B^2>A^2$, the central pressure is positive for $\Delta p_0>(B^2-A^2)/(8\pi\,A^2B^2)$.
Both $p_1(r)$ and $\Delta p(r)$ are monotonically decreasing, and therefore, $p_2(r)$ is non-negative and monotonically decreasing.
Since the pressure anisotropy is non-negative, $P_2(r)$ is positive.
See Figure~\ref{fig:Range_Second_Paper}.
If $P'_1(r)\leq -\Delta p'(r)$, then $P_2(r)$ is monotonically decreasing.
As can be seen from Figure~\ref{fig:Example_Second_Paper}, the choice $A=5$, $B=10$, and $r_{b,2}=1.5$, for example, yields $\Delta p_0=0.00135$ and the solution satisfies all physical conditions.

To summarize, the metric
\begin{equation}
    \label{Astro_exact}
    ds^2=-\frac{\left(1+\Delta p_0\,I(r)\right)^2}{\left(1-\frac{r^2}{B^2}\right)}dt^2+\frac{dr^2}{\left(1-\frac{r^2}{A^2}\right)\left(1-\frac{r^2}{B^2}\right)}+r^2(d\theta^2+\sin^2\theta\,d\phi^2),
\end{equation}
where $I(r)$ is given by (\ref{Ir}),
is a three-parameter family of exact solutions that satisfies all the desirable physical conditions, at least for certain ranges of parameters. It also salifies the more-stringent condition of monotonic decrease of sound-speed.\footnote{Of the $127$ exact solutions of the isotropic system studied in \cite{delgaty1998physical}, $16$ satisfy all the physical conditions mentioned earlier. Of these, $9$ satisfy the this additional condition.}

\section{Conclusion}
As anisotropic fluid solutions are becoming increasingly relevant in astrophysics and cosmology (see the introduction), it is important to understand the solution space of the system and to see how solutions of physical interest are contained within it. To this end, in \cite{Akbar:2020qrr} and in this paper, we systematically studied algorithms that generate all solutions of the system and maps that transform one solution into another. An overall feature of the anisotropic system that emerges from our study is that the question of central regularity can easily be incorporated with appropriate choices of variables. This also places one in position of strength to search for solutions that satisfy other physical conditions and led us to find a few examples. In addition, we found that the study of the algorithms and maps can be combined together in fruitful ways. The combination of \textbf{Theorem 4.1} in \cite{Akbar:2020qrr} and \textbf{Lemma 2} in this paper, which lead to a three-parameter exact solution satisfying all physical conditions, is probably the first of the many  applications that will consequently be explored within this framework. We also examined the possibility of converting the generalized TOV equation into a Bernoulli equation under different constraints which provide additional insights into the solution space. Overall, the anisotropic system is very compatible with physical conditions if one uses the right variables and starts with regularity at the center and may lead to surprising resolutions of outstanding puzzles in the future.

\section*{Acknowledgements}
We thank Tiberiu Harko for useful communications. RS acknowledges the support of Margie Renfrow Student Funds and a Julia Williams Van Ness Merit Scholarship.

\end{document}